\documentclass[a4paper]{jpconf}
\usepackage{graphicx}
\begin{document}
\title{Temperature dependence in interatomic potentials and an
  improved potential for Ti}

\author{G.J.Ackland}

\address{School of Physics,
JCMB,  The King's Buildings, 
University of Edinburgh,
EH9 3JZ.}

\ead{gjackland@ed.ac.uk}

\begin{abstract}
The process of deriving an interatomic potentials represents an
attempt to integrate out the electronic degrees of freedom from the
full quantum description of a condensed matter system. In practice it
is the derivatives of the interatomic potentials which are used in
molecular dynamics, as a model for the forces on a system.  These
forces should be the derivative of the free energy of the electronic
system, which includes contributions from the entropy of the
electronic states.  This free energy is weakly temperature dependent,
and although this can be safely neglected in many cases there are some
systems where the electronic entropy plays a significant role.  Here a
method is proposed to incorporate electronic entropy in the Sommerfeld
approximation into empirical potentials.  The method is applied as a
correction to an existing potential for titanium.  Thermal properties
of the new model are calculated, and a simple method for fixing the
melting point and solid-solid phase transition temperature for existing 
models fitted to zero temperature data is presented.
\end{abstract}

\section{Introduction}
Classical molecular dynamics simulations are widely used in
many areas of the physical sciences.  Eliminating the explicit
treatment of electronic degrees of freedom brings a computational
advantage of around six orders of magnitude in the simulation of
metallic systems, large enough to tackle systems which would be
crippled by finite size effects in an electronic structure
calculation.  For applications in microstructure, plasticity and
radiation damage where one wishes to study the behaviour of the atoms
rather than electronic structure, this makes classical MD the method
of choice.

The price for this is that the interatomic forces must be represented
by the derivative of an interatomic potential, which is necessarily a
poor approximation to the electronic structure.  Typically this
potential involves some functional form fitted to empirical or {\it ab
initio}-calculated data.  It is useful to distinguish the two distinct
sources of error involved in this process: incorrect functional form
and poor fitting.  

Many MD
simulations are still conducted using a pairwise-additive interatomic
potential, such as Lennard Jones or Coulomb charges.  Such pairwise
potentials constrain possible values of the elastic constants.  Most
notably, the ``Cauchy'' relation which relates $C_{12}$ with $C_{66}$.
In a pairwise potential, elastic constants are given by the second
derivative of the energy with respect to strain.  Regarding
the potential as a function of $r^2$ rather than
$r$, it is easy to show that for {\it any} pair potential:
\[ C_{12} = C_{66} = \frac{2}{\Omega}\sum_{ij} V''(r_{ij}^2) x_{ij}^2y_{ij}^2  \]
where $r_{ij},x_{ij},y_{ij}$ are the separation between atoms $i$ and
$j$, and x,y components thereof. where $i,j$ run over all atoms and
$\Omega$ is the volume of the system.  Experimentally, this is true 
only for strongly
ionic materials. Table 1 shows that elasticity in NaCl can be
described by pairwise forces, but it is impossible to fit the three
independent elastic constants for typical oxides, noble gases, water
and metals. No amount of fitting can circumvent this mathematical
constraint: the problem is one of of incorrect functional form.

\begin{table}[h]
\caption{\label{tabone}Elastic constants for selected materials
 showing the violation of the cauchy relation $C_{12}=C_{66}$ and by
 implication the impossibility of describing these materials by a
 pairwise additive potential. For water the constants are $C_{13}$
and $C_{44}$, the hexagonal equivalent relation}
\begin{center}
\lineup
\begin{tabular}{*{5}{l}}
\br                              
Material& $C_{11}$ &$C_{12}$&$C_{66}$\cr 
NaCl & 482 &128 &127\\  
MgO  & 291  & 96&152\\
Argon  & 233 &149 &117\\
SiC  & 385 & 135 & 257\\
Cu  & 168 &121 &75  \\
Fe    &237 &141 &116\\
Water ice&  & 59  & 31 \\
\br
\end{tabular}
\end{center}
\end{table}
There are many similar examples where an incorrect functional form
renders the process of fitting impossible.  About 25 years ago there
was a strong move away from pairwise forces, in particular in metals,
in favour of models which describe the local environment such as the
Embedded atom method\cite{eam}, Finnis-Sinclair\cite{fs}, and Tersoff\cite{tersoff}.  Easily justified by
elementary consideration of the electronic structure, these functional
forms allowed enough freedom in fitting to avoid the Cauchy constraint
and similar relations governing surface and vacancy energies.  They
satisfy the needs of most applications, and even if transferrability
is imperfect have remained the workhorse of atomistic simulation ever
since.

Some more recent potential developments have simplified the fitting
process or made it more intuitive.  The Modified Embedded Atom Method\cite{meam}
maps local coordination onto spherical harmonics and hence establishes
a link to electron orbitals.  Two band and magnetic potentials\cite{reed,dd,jnm}
explicitly introduce limited electronic degrees of freedom, and then
show how to eliminate them analytically, leaving a simple potential of
embedded atom form, but justifying more complicated parameterisations.

Ultimately, however, derivation of a functional form for a potential
should go back to first principles.  The overwhelming success of the
Kohn-Sham formulation of the density functional theory suggests that
we should start with the Kohn-Sham Hamiltonian\cite{kohn}, which in standard notation is:
\begin{equation} 
F(\rho) = T[\rho] + \frac{1}{2}\int\frac{\rho({\bf r})\rho({\bf r}')}
{4\pi\epsilon_0|{\bf r-r'}|} d^3{\bf r} d^3{\bf r'}
+ E_{xc}[\rho] + \sum_i\int\frac{Z_ie\rho({\bf r}')}{4\pi\epsilon_0|{\bf R_i-r'}|} d^3{\bf r'} 
+ \sum_i\sum_j\frac{Z_iZ_je^2}{4\pi\epsilon_0|{\bf R_i-R_j}|}  \label{ks} \end{equation}
For molecular dynamics we are interested in forces.  These can be
calculated using the Hellmann-Feynman theorem\cite{hf} which tells us
that the forces are simply the {\it partial} derivative of the
Kohn-Sham Hamiltonian with respect to ionic positions {\bf $R_i$}.  Inspection of
eq.\ref{ks} shows that only the final two, electrostatic, terms
contribute to this.

At finite temperature the electron free energy for a metal changes as
the Fermi distribution becomes non-singular.  The energy increases as
states above the Fermi are occupied, while the electronic entropy also
increases due to partial occupations.  According to Sommerfeld theory\cite{AM}
there is an additional temperature-dependent contribution to the free
energy which depends on temperature and the density of states at the
Fermi energy as
\[ F_{som}(T) \propto T^2 n(E_F) \] 

In many electronic structure packages this effect is exploited to
improve the numerical stability of the self consistency loop, by
calculating with finite temperature electrons, so-called ``Fermi
Smearing''.  Typical effective temperatures can run to thousands of
Kelvin, and the result is adjusted back to zero Kelvin.  Nevertheless,
the Sommerfeld temperature dependence of the electronic free energy is real, and
so any potential involving integrating out electronic
degrees of freedom should itself be temperature dependent.  It is
worth recalling at this point that the Hellmann-Feynman theorem is
valid only for an electronic energy which is a {\it variational
  minimum} with respect to the parameters describing the electronic
structure.  For the Fermi-smeared system the
variational quantity is the electronic {\it free energy}
($U_{el}-TS_{el}$), so the Hellman-Feynman forces are the derivative
of this.  The physical assumption underlying this is similar, though
not identical, to the Born-Oppenheimer approximation: the ionic motion
should be slow enough for the electrons to relax into their
equilibrium distribution.

\begin{figure}[h]
\hspace{1cm}\begin{minipage}{18pc}
\includegraphics[width=18pc]{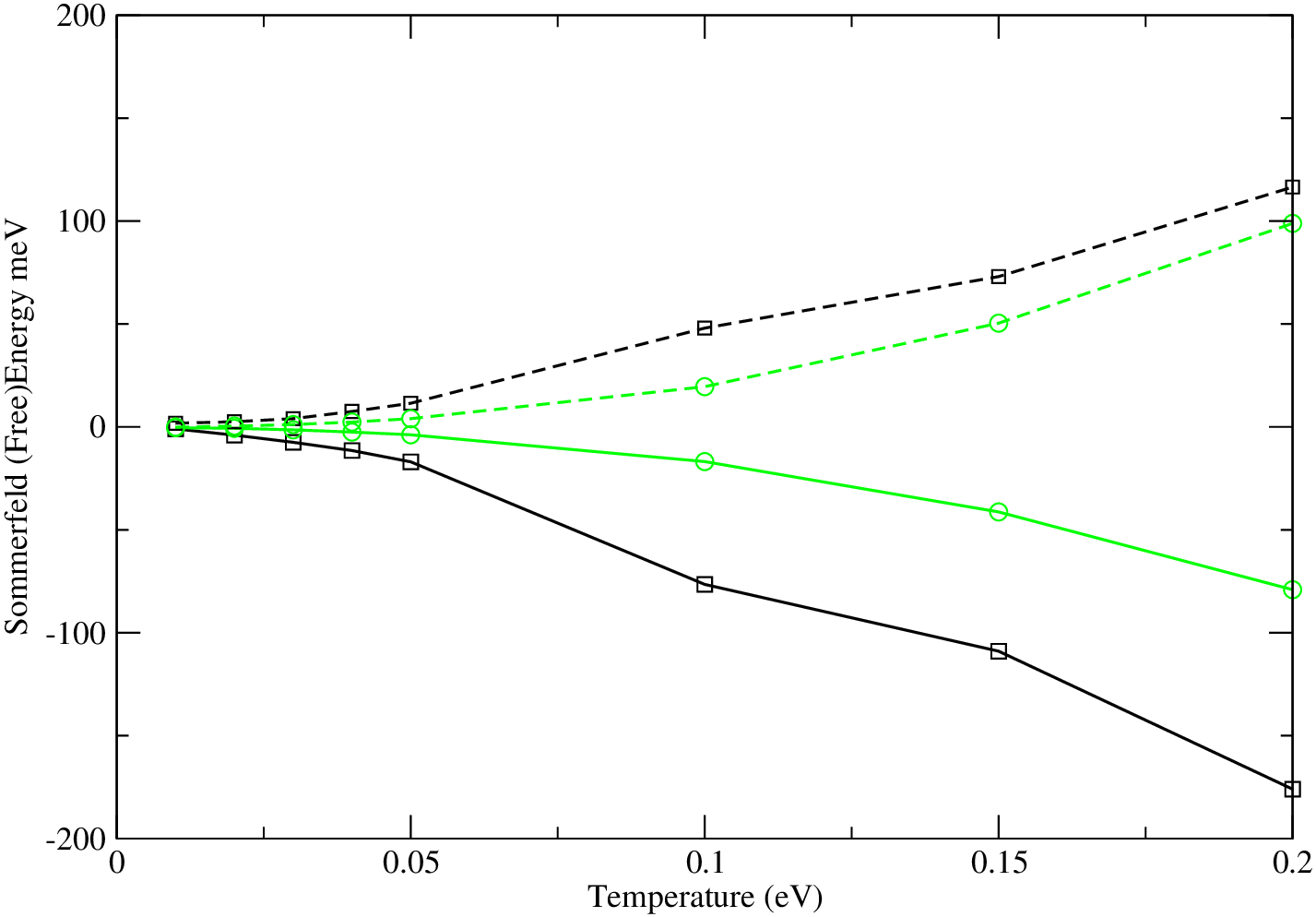}
\caption{\label{fig:TiSom}Sommerfeld electronic energy (dashed lines) and 
free energy (solid lines) for bcc (black, squares) and hcp (green, circles) Ti}
\end{minipage}\hspace{2pc}%
\begin{minipage}{18pc}
\includegraphics[width=18pc]{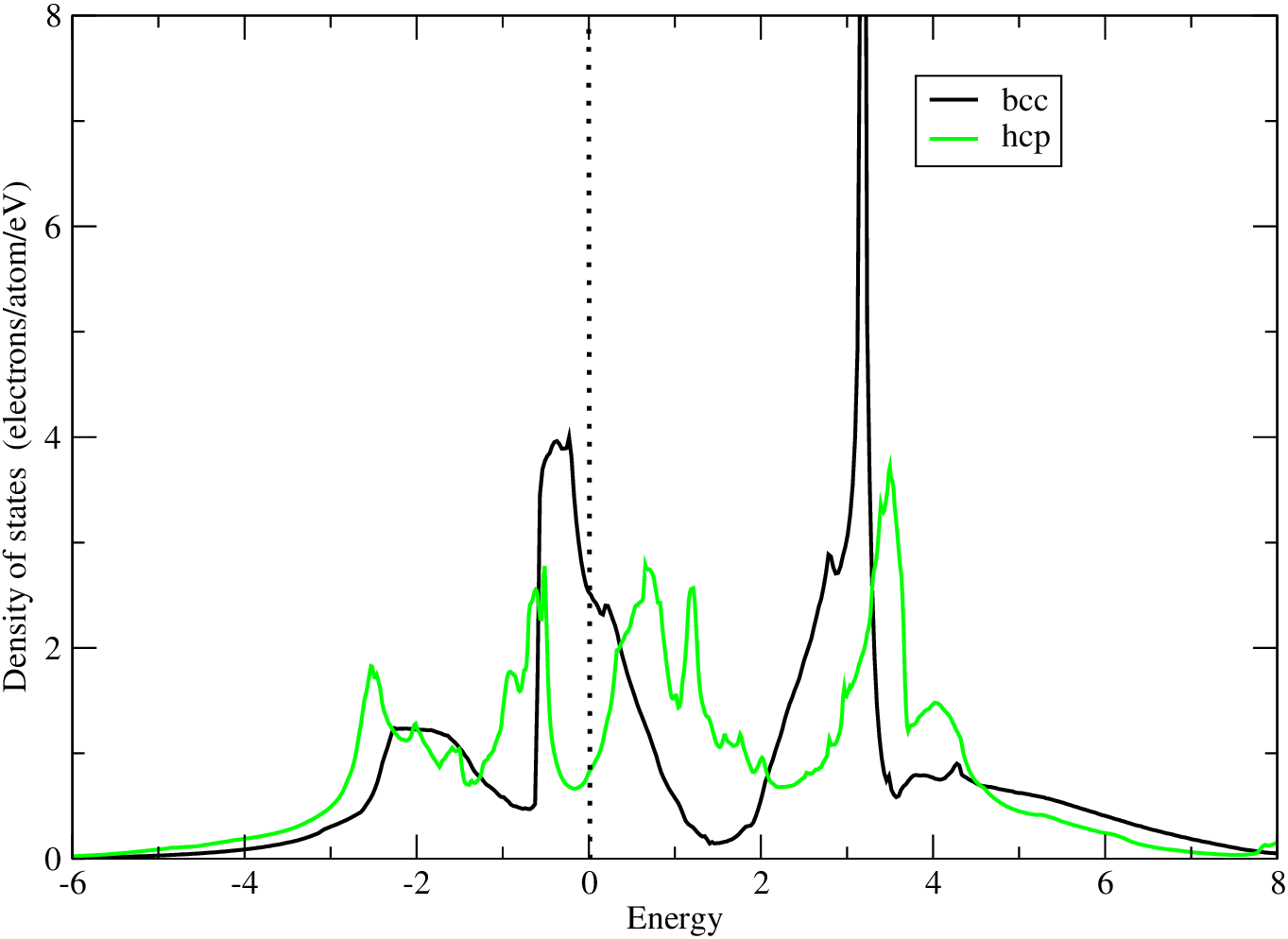}
\caption{\label{fig:TiDOS}Density of states for bcc (black) and hcp (green) Ti}
\end{minipage} 
\end{figure}

The temperature dependence of the Sommerfeld electronic free energy
for Ti is shown in Fig.\ref{fig:TiSom} alongside the associated
density of states.  
These figures were calculated using the
CASTEP\cite{castep} program with settings as given in previous
work\cite{tegner}, using Fermi-Dirac smearing with the smearing width
corresponding to the quoted temperature.  In all cases the atoms were
located on their ideal lattice sites.  The density of states was
calculated using the {\it castepdos} program, which interpolates and
integrates the density of states using the octahedron method between
the explicitly calculated k-points.\cite{zrh}
At 0.05eV (about 600K) the Sommerfeld contribution in bcc is
34meV compared with 7meV for hcp.  This is about 20\% of the free
energy difference at 0K.  

\section{Sommerfeld Potentials}

In order to include Sommerfeld effects in 
molecular dynamics,
the potential must be written in the form of a sum of
energies per atom $j$ in an arrangement of atoms $i$ in positions 
$\{{\bf r_i}\}$ :
\begin{equation} U(\{{\bf r_i}\}) =  \sum_j U_0({\bf r_j}, \{{\bf r_i}\}) + 
F_{som}({\bf r_j}, {\{\bf r_i}\})\label{U}\end{equation} The vast majority of interatomic
potentials are fitted to (extrapolated) zero-temperature experimental
data and/or {\it ab initio} data calculated on the Born-Oppenheimer
surface.  Thus in many cases good parameterisations for $U_0$ already
exist.  Here we consider the second term in equation \ref{U}, which  is proportional
to $T^2$. The
proportionality constant can be treated as a fitting parameter ($A_T$).
The challenge is to obtain a sensible measure for the local density of
states at the Fermi energy $n_j(E_f)$ in terms of the atomic positions
without performing a full electronic structure calculation.  The most
convenient approach is to write it as a pairwise function:
\[n_j(E_f)= \sum_i f(|{\bf r_i- r_j}|)\]
We could stop at this point, and simply take $f$ as an arbitrary
function to be fitted, however it is worth discussing how it might
look.
\subsection{Heuristics for f(r) in transition metals}
$f(r)$ attempts to measure the density of states at the Fermi level
projected onto an individual atom.  This quantity can be readily
calculated for various crystal structures (e.g. Fig 2), and may vary quite widely
between them.  If we consider a canonical d-band model, the density of
states is very different between close-packed and bcc structures.  The
DoS can be defined in terms of moments, in particular the third and
fourth moment determine the skew and kurtosis.  The moments theorem\cite{cyrot}
relates these to a real-space picture involving closed paths of near
neighbour hops.  This defines the key difference between close
packing (many loops of three hops) and body centred cubic (no three-hop
loops- many four loops).  Open structures seldom feature in transition
metals and their alloys, so we can neglect these. 

Although counting paths of hops cannot be done directly by a pair
potential, it suggests a related local measure. For practical purposes
the skew/kurtosis relation can be approximated by using a
function which is either sharply peaked around first neighbour
(favours three-loop equilateral triangles) or a broader function which 
favours four-loops.  Which is energetically favoured for a given material 
depends on the band filling.

In most existing EAM and Finnis Sinclair potentials this narrow vs
broad minimum in the potential determines the stability of close
packed vs bcc materials.  In the moments picture, the longer range
acts as a proxy for measuring four-membered rings.  Use of exponential
fitting functions, as in early EAMs, tends to give a narrow minimum;
polynomial functions, as used in Finnis Sinclair potentials, give a
broader minimum.  This, rather than any deep physics, explains their
early application to fcc and bcc respectively.

\subsection{Application to Titanium}
For titanium the Fermi level falls at a minimum of the hcp density
of states, and a maximium of the bcc density of states.  Thus we can
expect that electronic entropy contributions will be different and
significant in determining the phase transition temperature.  Indeed,
electronic entropy was shown
to provide
almost half the excess entropy of bcc Zr\cite{Willaime}, a material very similar to Ti.
The density of states is for valence electrons, so $f(r)$ should have
significant value only outside the repulsive core.  It is also desirable
that its derivatives be continuous and the number of adjustable
parameters is minimised. Consequently we write the function as:
\[ f(r) = X^2(1-X)^2   \hspace{0.5in} 0<X<1; \hspace{1in} X=(r-r_o)/d\]
\[ F_{som} = \sum_{ij} A_T T^2 X^2(1-X)^2 \]
this gives a three parameter model, defining the range of the
interaction from $r_0$ to $r_o+d$ and the strength $A_T$

Most published potentials are fitted to zero temperature data, so
this Sommerfeld term can simply be added.  Moreover, the parameterisation is
relatively straightforward: we have calculated the
temperature-dependent contribution to the free energy for fcc and hcp
Ti.  These two pieces of data are sufficient to parameterise 
the model.

For Ti we consider a potential introduced in 1992\cite{Ti} which is
known to have too low a basal stacking fault but otherwise reproduces
the behaviour of hcp titanium reasonably well.  From Fig. \ref{fig:TiSom}
above we see that at low energies (up to 1000K) the Sommerfeld
contribution to bcc is about 4.7 times that of hcp.  This suggests
an approximate solution $r_o=2.84\AA$, $d=1.46\AA$, $A_T=-7.5\times 10^{-7} meV/K^2$.

\subsection{Which temperature to use?}

In order to use the temperature-dependent potential, it is necessary
to specify the electronic temperature. In practice, most MD
simulations are relatively small and one can assume that the electrons
are able to reach thermal equilibrium throughout the region. 


Thus for an NVT/NPT calculation run using a thermostat, one can simply
use the thermostat temperature to fix a constant potential throughout
the run.  For NVE one can use the instantaneous global temperature of
the simulation for all atoms, so although the potential may vary in 
time, it is constant across space. 

\section{Tests of the new potentials}

To test the thermal effects we have carried out melting point
calculations from bcc and hcp Ti, and thermal expansion calculations.

\subsection{Melting Point Calculations}

We determined melting point  using the coexistence method\cite{Morris}, 
using the
MOLDY program in the NPE ensemble\cite{moldy}.  By using intermediate sized
systems, 16000 and 13824 atoms for hcp and bcc respectively,  
 it is possible to suppress the bcc-hcp transition and
calculate melting points from both hcp and bcc phases.
The original
potential\cite{Ti} gives a melting point of $1395\pm10$K (hcp) and $1790\pm10$ K (bcc),
the difference indicating the stability of bcc at high temperature.
This contrasts with the 0K energies of -4.853eV (hcp) and -4.807eV
(bcc) which indicate low temperature stability of hcp.  The transition
at intermediate temperature is via a soft phonon analogous to
zirconium\cite{Willaime,pinsook}, and care must be taken to ensure that the solid phase does not transform during the simulation, which we did by monitoring local coordination using the BallViewer code\cite{ballviewer}.

Experimentally, the melting temperature of Ti is 1941K, and the
bcc-hcp transition is at 1155 K.  The original potential therefore
gives too low a melting temperature.  The effect of the Sommerfeld
correction on the melting is small (1315K and 1800K), slightly
stabilising bcc and destabilising hcp.

\begin{figure}[h]
\hspace{1cm}
\includegraphics[width=18pc]{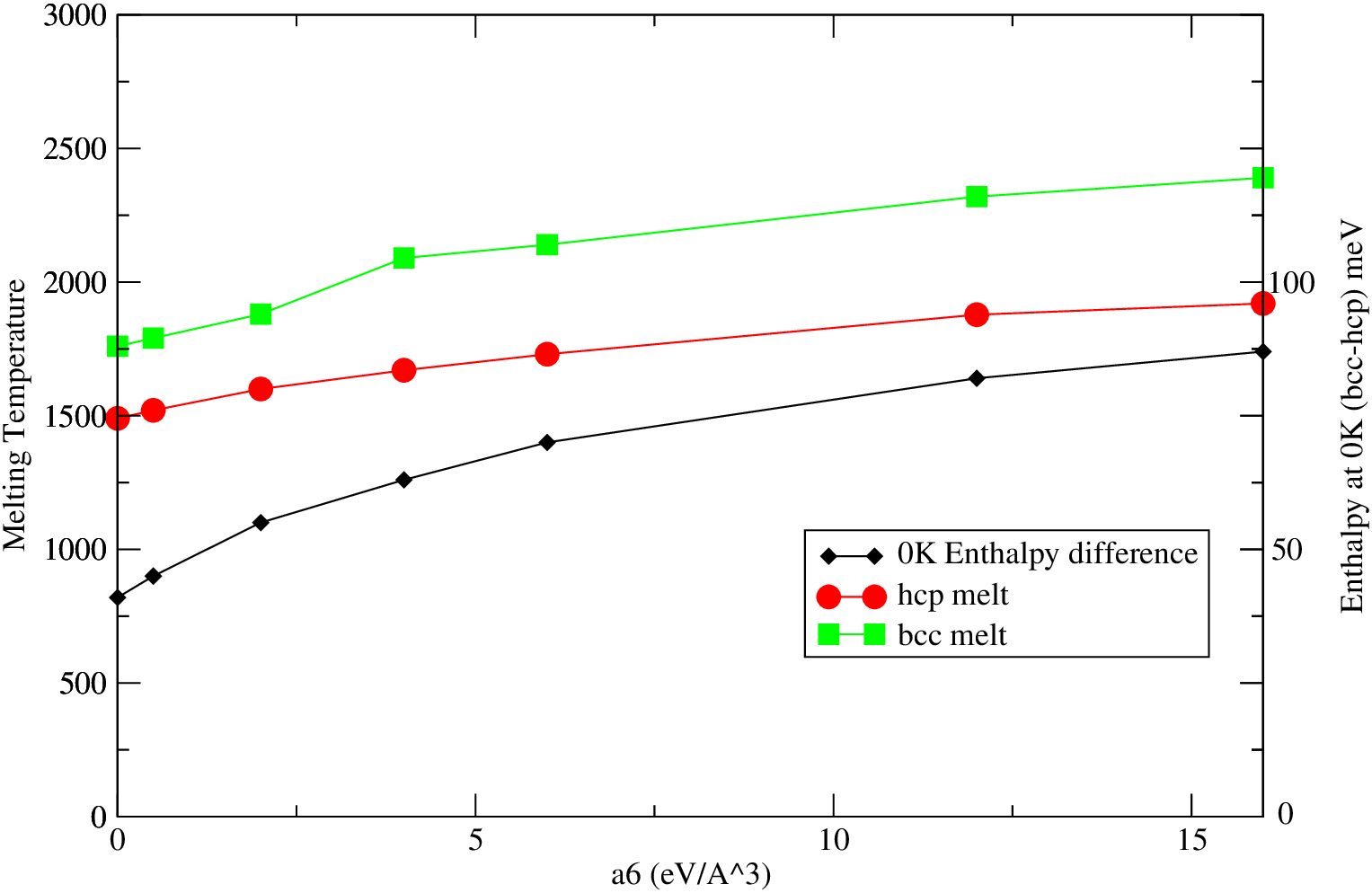}
\caption{Effect of short ranged repulsion of melting
  point and zero-temperature hcp-bcc stability.  See Appendix A for
  definition of $a_6$ \label{fig:melt}}
\end{figure}

Metals melt at high temperature because the liquid phase has higher
entropy than the solid.  Thus the more of the phase space the system
can reach without too high potential energy, the lower will be the
melting point.  Since a melt samples interatomic distances shorter
than typically realised in the solid, softening the potential inside
nearest neighbour distances can be expected to favour the liquid.  In the present potential, the $a_6$ spline parameter controls the short range repulsion.
The 1992
potential was not fitted to any data for short-ranged interactions, and
in Fig. \ref{fig:melt} we show the evolution of the melting
temperature as a function of the parameter $a_6$: the short ranged
spline starting close to nearest neighbour separations (see Appendix A). 
 The effect on
the melting point is very pronounced, while the 0K properties are
essentially unaffected.


The original potential had $a_6=0.494$eV/\AA$^3$:
the experimental melting point (1941K) and ab initio hcp-bcc
difference (90meV) were not fitted in 1992, and to do so requires a
significantly stiffer short-ranged potential. 
\subsection{Thermal Expansion}
The thermal expansion is not typically fitted in empirical potentials,
and in consequence is often significantly wrong.  The Sommerfeld
correction introduces additional temperature-dependent anharmonicity
into the potential, which affects the thermal expansion.  For this
potential the thermal expansion is close to the experimental data but
highly temperature dependent, especially at high temperature where 
the softening around second neighbours leads to high values.

\subsection{Phase Transformation Calculations}
 Whatever value is used,
the Sommerfeld correction will be insufficient to reduce the enthalpy
of bcc below hcp: the bcc structure is always stabilised by phonon
entropy. The constraints imposed by the crystal symmetry mean that it is not
possible to calculate the phase transition temperature between hcp and
bcc by coexistence.  Rather, it requires a full free energy
calculation\cite{free,caro} which can be performed by integrating using the
Gibbs-Helmholtz equation:
\[  \Delta (\frac{G}{T}) = -\int H/T^2 dT  \]
The right hand side of this equation can be calculated from a
single-phase NPT molecular dynamics calculation (see e.g. Appendix B), varying
the temperature slowly by incrementing the required temperature of a
Nose thermostat.  In practice we use this equation with the two
melting points and assume that $H/T^2$ varies linearly with $T$. The
stabilising effect on bcc lowers the transition temperature
considerably, the actual value depending on the chosen value of $a_6$.

\begin{figure}[h]
\hspace{1cm}
\includegraphics[width=18pc]{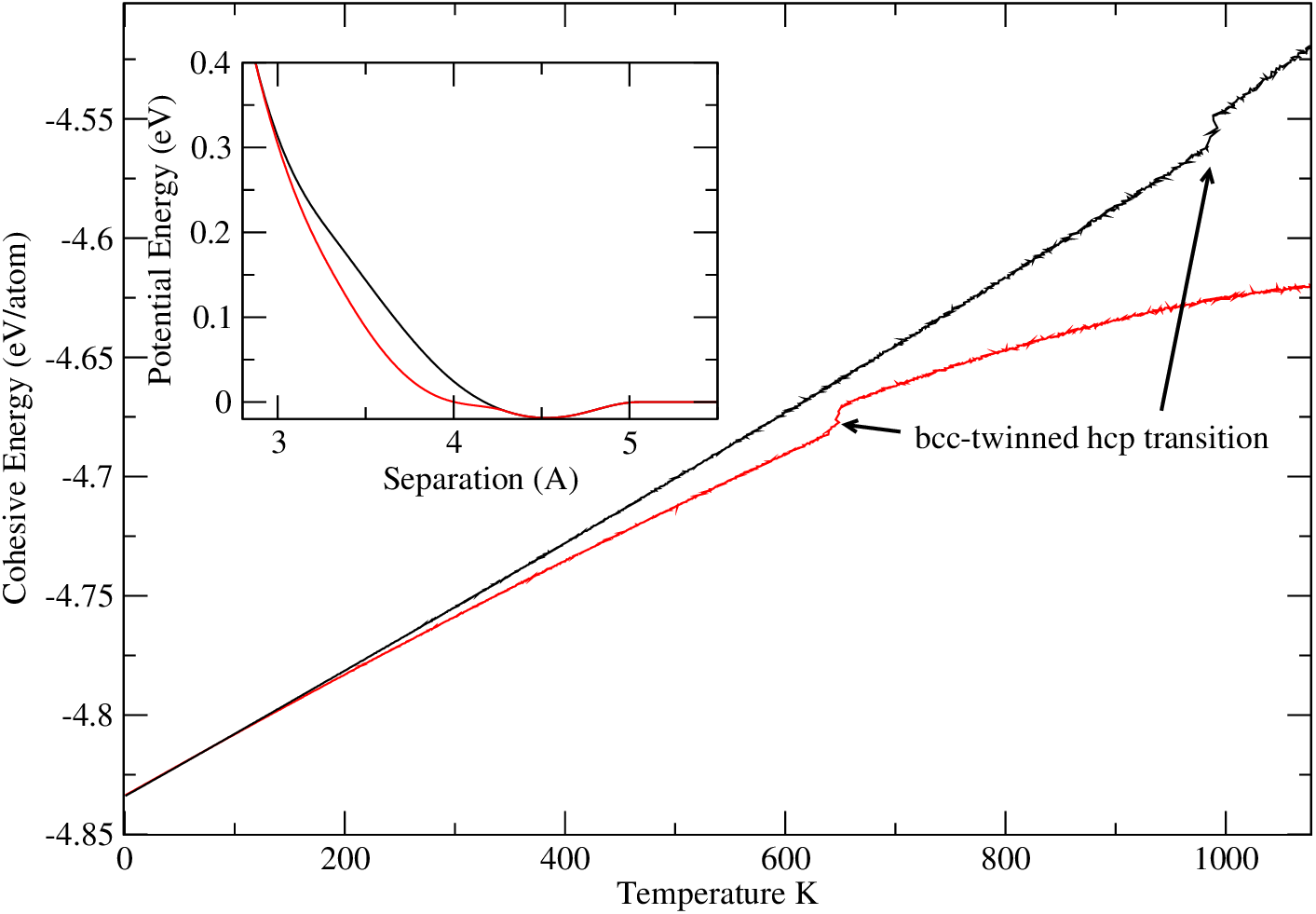}
\caption{Plot of cohesive energy per atom vs temperature for MD
  simulation on heating with and without Sommerfeld correction.  Main
  figure shows linear increase in energy with temperature, as expected
  from virial theorem (original potential, black line).  Corrected
  potential (red) shows deviation from virial behaviour as electronic
  degrees of freedom are excited at high T.  Both cases have a
  discontinuity as the cooling bcc transforms to twinned hcp, the
  correction stabilises bcc, dropping the observed transformation from 1000K to 650K.  Inset shows pair potential V(r) at 0K and 1250K \label{coh}.}
\end{figure}

Figure \ref{coh} shows the effect of the Sommerfeld potential in a
dynamics calculation of the phase transition.  Although this is only 
a lower bound for the thermodynamic transition temperature, 
it does illustrate the increased stability range of bcc.

\section{Conclusions}

The thermal excitation of electronic degrees of freedom introduces a
temperature dependence into the electronic free energy.  If the
electronic degrees of freedom are integrated out to make an
interatomic potential, this temperature dependence should remain.
Where the excitation arises simply from the Fermi-Dirac distribution
via Sommerfeld theory, this temperature dependence scales as $T^2$.
The magnitude of the effect has been calculated for Ti using density
functional theory, and found to give a contribution of a few tens of
meV to the bcc-hcp free energy difference.  This in turn is enough to
lower the calculated hcp-bcc phase transition temperature
significantly, but has less effect on the melting temperature.
Neither effect is large compared to the errors of typical EAM-type
potentials which do not include melting temperature as a fit
parameter.  

By contrast, the melting temperature has strong correlation with the
short ranged part of the potential, inside the normal separations
found and fitted at 0K.  By adjusting only the short-range terms it is
possible to refit the melting point of a potential fitted to elastic
moduli, cohesive energy, vacancy and surface formation etc. without
disturbing those properties. 
This suggests a simple way to improve the
high temperature phase performance of existing potentials
without major redevelopment\cite{Pasianot,bonny,winey}. Note that
energies of self interstitials\cite{fs,eam,dd} and interstitial
impurities\cite{zrh,fec} may be affected.

In addition to Sommerfeld correction, a strong temperature dependence
of the electron free energy may be found in magnetic materials such as
iron\cite{has1,has2,ma}, where the fcc phase is stabilised by
paramagnetic free energy.  This effect could be captured by a
temperature-dependent potential, although the excitation of all
magnetic modes means the temperature dependence would be more complex,
requiring an additional many-body term rather than a pairwise one.
This will be the subject of future work.

\section{Appendix A}

The potential has a functional form for the energy of the $i^{th}$ atom 

\begin{eqnarray} U_i& = &\sum_j\sum_k H(r_k-R_{ij}) a_k(r_k-R_{ij})^3  \\
 & & \sum_{j} A_T T^2 X^2(1-X)^2H(r_o+d-R_{ij})H(R_{ij}-r_o)\\
 & & - \sqrt{\sum_j\sum_k H(R_k-R_{ij})A_k(R_k-R_{ij})^3} 
\end{eqnarray}
with $X$ as defined above, $R_{ij}$ the separation between $i^{th}$ and  $j^{th}$
nearest atoms, $H$ the Heaviside step function and the other parameters given inTable A1 below.

\begin{table}[h]
\begin{tabular}{c|cccccc}
\hline
$k$  & 1 & 2 &3 & 4 & 5& 6 \\
\hline
$a_k$ & -0.785715& 1.110966 &-0.299450 & -0.143061 &1.025368& 0.494293\\
$r_k$ &5.09113& 5.00767& 4.673828& 3.964408& 3.338449& 2.9508\\
$A_k$ & 0.547614& -0.551266      \\
$R_k$ & 5.09113& 4.381714        \\
\hline
&  $A_T=-7.5{\small \times}10^{-7}$& $r_o$=2.84& $d$=1.4\\ 
\hline\end{tabular}
\caption{Table A1. Parameters for the Ti potential, from ref.\cite{Ti}, converted into units of eV and Angstroms, and for the temperature-dependent correction.  Note that $a_6$ can be varied to fit the melting temperature.}
\end{table}
\begin{figure}[h]
\hspace{1cm}
\includegraphics[width=18pc]{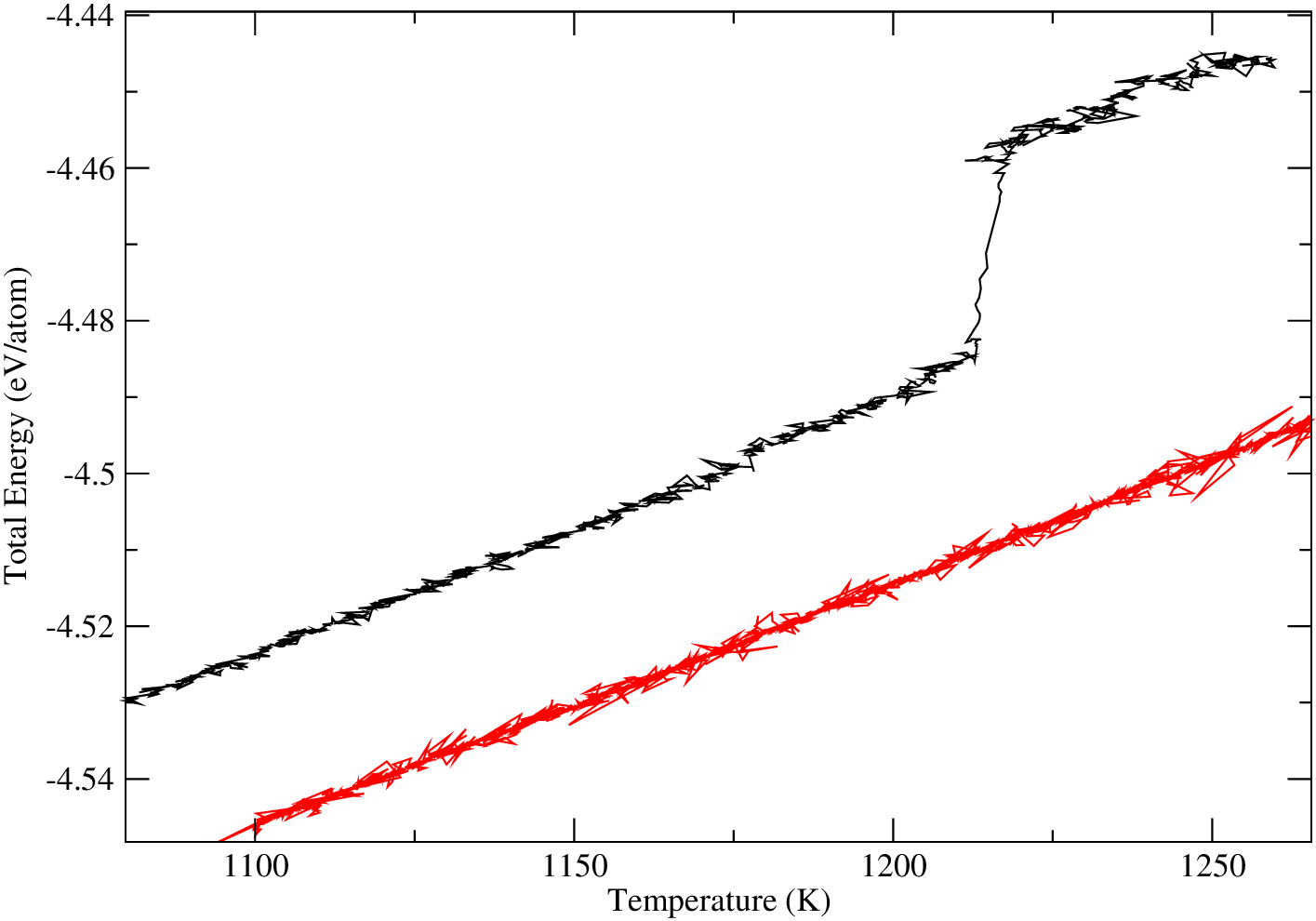}
\caption{Plot of enthalpy vs temperature showing failure of
  thermodynamic integration due to phase transition from bcc to
  deformed hcp on cooling.  Upper line (black) is bcc at high
  temperature, transforming to twinned hcp on cooling.  Reheating 
  gave hysteresis in the retransformation.  Note the
  different transuition point compared with fig 4. Lower line (red) is
  hcp on heating from 0K: recooling produced the same curve.  
  NPT simulations were run with a cooling rate of 1K/ps and a
  1fs timestep, with a Nose relaxation time parameter of 100fs.
  Phases were initialised in a supercell with periodic boundaries
  compatible with the phase in question: although the transformation
  could happen in principle in Parrinello-Rahman dynamics, in practice
  the transformation is slow enough to gather phonon
  statistics.\label{Z}}
\end{figure}

\section{Appendix B}
One potential drawback of the thermodynamic integration is that the
system will undergo a phase transformation as the temperature is
varied, rather than remaining in the metastable phase.  When this
occurs, the latent heat causes a discontinuity in a graph of enthalpy
vs temperature.  Fig.\ref{Z}. shows an example of this in cooling of the bcc
phase. Some authors have maintained that if the transformation occurs
quickly enough, the temperature of the discontinuity would be the
transformation temperature.  However, better statistics than are
available here are needed to ensure that this is so\cite{Zalf}.
Moreover, Fig \ref{Z}.  shows that the transformation is not to the pure hcp
phase, and BallViewer analysis\cite{ballviewer} shows that the low
temperature phase is a twinned hcp.

\end{document}